# A Symbolic-Numerical Approach for the Sensitivity Analysis of Dielectric Resonator Sensors

**Remi Barrere, Pierre Boughedaoui, Michelle Valentin**

University of Franche-Comte, ENSMM
26 chemin de l'Epitaphe
F-25000 Besancon, France
rbarrere@ens2m.fr

A theoretical model based on the transverse resonance method is proposed for the description of cylindrical multilayer dielectric resonator sensors. From this model, the resonant frequency and the sensitivities with respect to geometrical and physical parameters are computed by means of a combination of symbolic and numerical procedures. These are gathered together in a package in view of the computer assisted design of this range of sensors. On this occasion, a few design patterns for engineering applications are sketched.

February 12, 2003

## 1. Introduction

**Dielectric resonator sensors**

In recent years, considerable attention has been devoted to dielectric resonators with a high permittivity, a low material loss (hence high quality factor Q) and a very low resonant frequency temperature coefficient. These compact dielectric resonators alternate the conventional metallic resonant cavities at microwave frequencies. Figure 1 shows the field distribution of the lowest order mode $TE_{01\delta}$ : unlike a metallic resonator, the field extends beyond the bulk of the device. This evanescent field enables external coupling of microwave energy in the vicinity of the dielectric resonator.

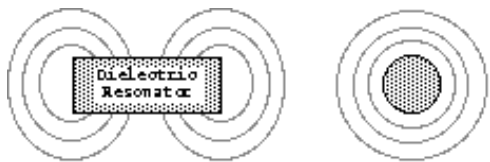

**Figure 1.** Field distribution of the TE01$\delta$ mode of a dielectric resonator : magnetic field lines are shown on the left, whereas electric field lines are shown on the right.

In applications such as low noise sources and low loss filters, the evanescent nature of the field patterns requires that the resonator is screened by a metallic housing to avoid radiation losses and environmental influences. In microwave sensor applications, dielectric resonators are best suited because of the environmental influence on the resonant frequency and the high accuracy of frequency measurements.

**Modelling and computing**

The theoretical models of electromagnetic devices generally lead to equations that have no known exact solution, except for the case of simple structures. That is why physicists or engineers most often resort to numerical techniques, such as the finite difference method [1] or the finite element method [2]. However, these have drawbacks, especially in the case of sensitivity analysis; indeed, purely numerical techniques entail repetitive computations where a single symbolic formula would give the required result.

These are the reasons why a mixed symbolic-numerical approach is put forward. It is based on the transverse resonance method, which leads to a system of transcendental equations. The computation of resonant frequencies, which entails the solution of that system, rests on a numerical technique: either the secant method or Newton's method. Then, the computation of sensitivities, which consists in evaluating partial derivatives, rests on computer algebra capabilities.

**A package for the CAD of dielectric resonator sensors**

According to the technical application, the designer will choose a convenient set of geometrical and physical parameters and compute the resonant frequency and certain sensitivities. In order to optimize this choice, he or she may have to vary some of these parameters. A package called SensorDesign was created with that purpose, which can thus be used as a computer assisted design system for this range of sensors. Its core consists of an expression (a data structure) that describes the sensor with its parameters, together with the aforementioned numerical and symbolic procedures.

## 2. Mathematical model of the sensor

**Description of the sensor**

The resonator consists of three layers, as shown in Figure 2:
• a substrate (dielectric cylinder with radius $b$, height $h_1$ and relative permittivity $\varepsilon_1$);
• a median layer with a core (cylindrical dielectric resonator with radius $a$, height $h_2$ and relative permittivity $\varepsilon_{dr}$) and a ring (with inner radius $a$, outer radius $b$, height $h_2$ and relative permittivity $\varepsilon_2$);
• an upper layer (dielectric cylinder with radius $b$, height $h_3$ and relative permittivity $\varepsilon_3$).

This multilayer structure is placed into a cylindrical metallic cavity of radius $b$ and height $h_1 + h_2 + h_3$, which acts as a shield. So the structure can be viewed as three cylindrical waveguides placed ends on, a heterogeneous waveguide being placed between two homogeneous ones. In a number of applications, $\varepsilon_2$ and $\varepsilon_3$ are identical. In all cases, the materials are supposed to be homogeneous and isotropic.

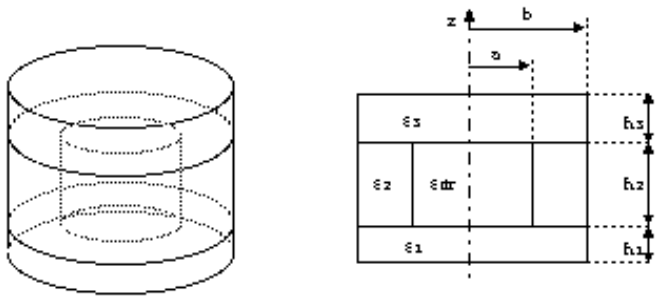

**Figure 2.** Geometrical structure and physical parameters of the resonator.

A variation of the resonant frequency occurs when $h_3$ or $\varepsilon_3$ varies, i.e., the frequency is a function of the displacement of the upper plate or of the material of the upper layer. So, the system can be viewed as a transducer that measures these displacements or permittivities, or other characteristics of the layers or the structure. Multiple applications are possible, such as displacement or pressure sensors, humidity probes or devices for measuring complex electromagnetic parameters.

**Modelling the sensor**

In order to make the model simple enough, a few physical assumptions are necessary. Conductors and dielectrics are supposed to be lossless. So, lossless propagation along the z-axis is assumed for the median heterogeneous waveguide, with evanescent waves in the ring. Evanescent waves are supposed to take place in the homogeneous waveguides at each end of the sensor. So, the core of the sensor is the resonant part of the structure. This is obtained by choosing the highest permittivity for it.

Only the first transverse electric ($TE_{01\delta}$) or magnetic ($TM_{01\delta}$) modes with cylindrical symmetry (no angular variation) will be considered. Experiments show that these modes are well isolated from higher order modes, thus avoiding undesirable couplings.

As shown in Figure 3, the volume of the sensor can be divided into 6 zones (3 cylinders $C_1$, $C_2$, $C_3$ and 3 rings $R_1$, $R_2$, $R_3$) with 3 surfaces of discontinuity: the planes $S_1$ and $S_2$ and the cylinder $S_3$ ($S_0$ refers to the ground). The electromagnetic field is supposed to propagate in the core $C_2$ and to be evanescent inside the cylinders $C_1$ and $C_3$ and the ring $R_2$. It is supposed to be zero inside the rings $R_1$ and $R_3$, except when b/a≃1.

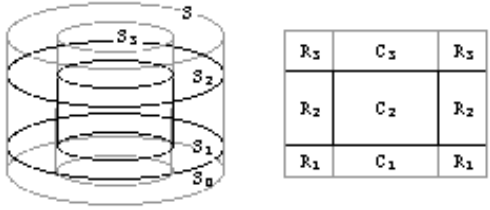

**Figure 3.** Three surfaces delimiting six zones, where different behaviors occur.

The choice of the model mainly depends on the ratio b/a. In view of comparisons with experimental data, when the ratio b/a is close to 1 (b/a≈1), the mediums 1 and 3 are assumed to be cylindrical electric-wall waveguides with radius b. Otherwise, when b/a is large (b/a≫1), these are assumed to be cylindrical magnetic-wall waveguides with radius a.

From these assumptions, a model of the resonator can be established, based on the transmission line analogy. Each layer is modeled by a transmission line called $TL_i$ (Figure 4) the characteristics of which are determined below.

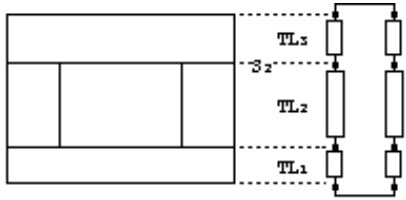

**Figure 4.** The multilayer sensor as a transmission line.

## Characteristic equation

The characteristic impedance $Z_{c_i}$ of the $i^{\text{th}}$ equivalent transmission line is expressed from the propagation constant $\gamma_i$ by: $Z_{c_i} = j\omega\mu_0/\gamma_i$ for a TE mode, and $Z_{c_i} = \gamma_i/j\omega\varepsilon_0\varepsilon_{r_i}$ for a TM mode. For lossless dielectric material, the permeability is $\mu_0$ and the permittivity $\varepsilon_0\varepsilon_{r_i}$ is real. The propagation constants are either imaginary or real, according to whether the modes are evanescent or propagate; they satisfy the following relationships, where $k_0$ is the wavenumber in vacuum:

$$\gamma_1 = \left(\left(\frac{x_m}{r}\right)^2 - k_0^2\,\varepsilon_1\right)^{\frac{1}{2}};\ \gamma_3 = \left(\left(\frac{x_m}{r}\right)^2 - k_0^2\,\varepsilon_3\right)^{\frac{1}{2}};\ \gamma_2 = j\left(k_0^2\,\varepsilon_{dr} - \xi_1^2\right)^{\frac{1}{2}}$$

The first two relations which concern homogeneous cylindrical waveguides are well-known [3]. For the median heterogeneous waveguide, the third relation is obtained with the assumption that propagation occurs along the rod internally and the fields decay exponentially outside the rod. Moreover, in that case, the characteristic equation connecting the radial wavenumbers in the dielectric media $C_2$ and $R_2$ respectively can be computed by means of the radial transverse resonance method [4, 5]:

$$\frac{J_0{'}(\xi_1\,a)}{\xi_1\,a} + \frac{P_0{'}(\xi_2\,a)}{\xi_2\,a} = 0 \text{ for a } TE_{0m} \text{ mode}$$

$$\varepsilon_{dr}\frac{J_0{'}(\xi_1\,a)}{\xi_1\,a} + \varepsilon\frac{R_0{'}(\xi_2\,a)}{\xi_2\,a} = 0 \text{ for a } TM_{0m} \text{ mode.}$$

where $P_0{'}(\xi_2\,a)$ and $R_0{'}(\xi_2\,a)$ derive from:

$$P_n(\xi_2\,r) = J_n(\xi_1\,a)\,\frac{K_n(\xi_2\,r)\,I_n{'}(\xi_2\,b) - I_n(\xi_2\,r)\,K_n{'}(\xi_2\,b)}{K_n(\xi_2\,a)\,I_n{'}(\xi_2\,b) - I_n(\xi_2\,a)\,K_n{'}(\xi_2\,b)}$$

$$R_n(\xi_2\, r) = J_n(\xi_1\, a)\, \frac{K_n(\xi_2\, r)\, I_n(\xi_2\, b) - I_n(\xi_2\, r)\, K_n(\xi_2\, b)}{K_n(\xi_2\, a)\, I_n'(\xi_2\, b) - I_n(\xi_2\, a)\, K_n(\xi_2\, b)}$$

- for the electric-wall model: $r = b$, $x_m$ is the $m^{th}$ zero of $J_n(x)$ for the $TE_{nm}$ modes and the $m^{th}$ zero of $J'_n(x)$ for the $TM_{nm}$ modes (where $J_n(x)$ denotes the $n^{th}$ Bessel function);
- for the magnetic-wall model: $r = a$, and the previous eigenvalues are still valid by inverting the TE and TM modes.

**The transverse resonance method**

The analysis of a multilayer structure by the transverse resonance method consists in determining an equivalent transmission line for the composite structure. The mode matching condition for the electromagnetic field is expressed in terms of impedance or admittance of the equivalent network on the planes of discontinuity.

The reference plane can be chosen anywhere, but for convenience, it is preferably chosen at a surface of discontinuity. For instance, in the following, $S_2$ will be selected as the reference plane. Then, $Z_+$ denotes the impedance of the short-ciuited transmission line $TL_3$, whereas $Z_-$ denotes the impedance of line $TL_2$ loaded by the shorted-circuited line $TL_1$ (Figure 4). The resonance condition of the system is obtained from: $Z_+ + Z_- = 0$.

These impedances are derived from the well-known formula of impedance transformation on lines [3]:

$$Z_+ = Z_{c_3}\, \text{th}(\gamma_3\, h_3)\ ;\ Z_- = Z_{c_2}\, \frac{Z_{c_1}\, \text{th}(\gamma_1\, h_1) + Z_{c_2}\, \text{th}(\gamma_2\, h_2)}{Z_{c_2} + Z_{c_1}\, \text{th}(\gamma_1\, h_1)\, \text{th}(\gamma_2\, h_2)}$$

So, the resonance condition $Z- + Z+ = 0$ can be rewritten:

$$\beta h = \arctan\left[\frac{\alpha_1}{\beta\, \text{th}(\alpha_1\, h_1)}\right] + \arctan\left[\frac{\alpha_3}{\beta\, \text{th}(\alpha_3\, h_3)}\right]$$

for $TE_{nmp}$ modes, or:

$$\beta h = \arctan\left[\frac{\varepsilon_{rd}}{\beta}\, \frac{\alpha_1}{\varepsilon_1}\, \text{th}(\alpha_1\, h_1)\right] + \arctan\left[\frac{\varepsilon_{rd}}{\beta}\, \frac{\alpha_3}{\varepsilon_3}\, \text{th}(\alpha_3\, h_3)\right]$$

for $TM_{nmp}$ modes.

Finally, the resonant frequency f is determined from the characteristic equation and the resonance condition, that is by solving a system of two transcendental equations.

**Computational aspects**

A numerical technique is required to solve this system, hence the name of the function : `NFrequency`. It directly uses `FindRoot`, i.e., the secant method or the Newton one. The former (the default one) turns out to be faster, whereas the latter may be used when precise results are required. The method is chosen by means of an option associated with `NFrequency`.

In the package, the initial values of the resonant frequency f and the eigenvalue x for the $TE_{01\delta}$ mode are respectively obtained by using the simplified physical model of the cylindrical dielectric resonator of which all surfaces are perfect magnetic-walls [6] and the approximate expression given by [7].

The following examples are processed with a 366MHz RISC CPU running Mathematica 4.0. `New[Sensor]` creates a sensor object with standard default values.

```
defaultSensor = New[Sensor]
```

```
Sensor[{h₁ → 0.00127, h₂ → 0.015, h₃ → 0.003,
  a → 0.0175, b → 0.025, εdr → 37., ε₁ → 10.5, ε₂ → 1., ε₃ → 1.}]
```

```
Timing[NFrequency[defaultSensor] (* Secant Method *)]
```

$\{0.116667 \, \text{Second}, \, 1.90227 \times 10^9\}$

```
Timing[NFrequency[defaultSensor, Method → NewtonMethod]]
```

$\{0.266667 \, \text{Second}, \, 1.90227 \times 10^9\}$

The values of the parameters are supposed to be given in the SI system and results are then given in the SI system without specifying their associated units. However, in view of the order of magnitude of dimensions and frequencies, it may be more practical to use multiples, such as millimeters or gigahertz. The conversion can be made by hand or with the help of the command `Convert` from the standard package `Miscellaneous`Units``, which is automatically loaded by `SensorDesign`. `Convert` yields the result with its associated unit. So there is in the package a complementary command `ConversionFactor` that gives the numerical conversion factor between two units.

```
NSensitivity[defaultSensor, h₃, Type → Absolute]
```

$-3.67881 \times 10^{10}$

```
Convert[% Hertz / Meter, Mega Hertz / (Milli Meter)]
```

$-\dfrac{36.7881 \, \text{Hertz Mega}}{\text{Meter Milli}}$

`Plot` should be used with care; indeed, the default number of points to be evaluated is 25, which may lead to excessive durations, for the function `NFrequency` is then evaluated 25 times. Consequently it may be advisable to decrease the value of `PlotPoints` and of `PlotDivision` to avoid the triggering of the adaptive sampling. This can be done inside each `Plot` command or at the beginning of the session with `SetOptions`. Fortunately, the resulting curves are generally smooth, so only a few points (7 points for instance) yield satisfactory results.

```
f[η_] := NFrequency[ChangeParameterValues[defaultSensor, h₃ → η / 10.³]] / 10⁹
Plot[f[η], {η, 0, 14}, PlotPoints → 7, PlotDivision → 1,
  AxesOrigin → {0, 1.75}, AxesLabel → {"h₃ (mm)", "f(GHz)"}]
```

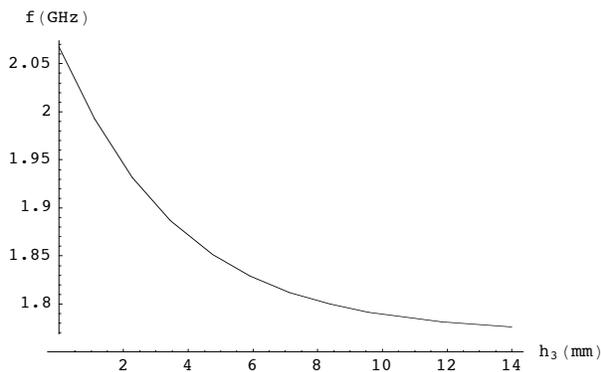

-Graphics-

## 3. Sensitivity analysis

### Mathematical aspects

Let's consider a sensor determined by n (geometrical or physical) parameters denoted $p_i$, so the resonant frequency f can be regarded as a function of these n parameters: $f(p_1, \ldots, p_n)$. Then, the absolute sensitivity $Sp_i$ (respectively relative sensitivity $RSp_i$, normalized sensitivity $NSp_i$) with respect to a specific parameter $p_i$ is given by:

$$\mathrm{Sp}_i = \frac{\partial f}{\partial p_i} \left( \text{respectively}: R\,\mathrm{Sp}_i = \frac{1}{f}\frac{\partial f}{\partial p_i}\,;\, \mathrm{NSp}_i = \frac{p_i}{f}\frac{\partial f}{\partial p_i} = \frac{\partial \ln f}{\partial \ln p_i} \right)$$

To compute the sensitivities, we must go back to the system of two equations in two unknowns f and x. In fact, the expressions also depend on the parameters $p_i$, so the system should be written:

$$\begin{cases} F_1(f, x, \ldots p_i, \ldots) = 0 \\ F_2(f, x, \ldots p_i, \ldots) = 0 \end{cases}$$

The solution $f(p_1, \ldots, p_n)$ is an implicit function; although no exact symbolic expression can be found for it, its partial derivatives can be computed in a symbolic way. Let $f_0$ and $x_0$ be the solutions to the system for a particular set of parameters $p_i = p_{i_0}$. Whether $f_0$ and $x_0$ are obtained by a numerical technique does not matter. Let us express the differentials of $F_1$ and $F_2$ at the "point $p_{i_0}$":

$$d F_k = 0 = \frac{\partial F_k}{\partial f} df + \frac{\partial F_k}{\partial x} dx + \Sigma_i \frac{\partial F_k}{\partial p_i} dp_i \quad k = 1, 2$$

Developing df and dx yields:

$$\forall i, \quad \frac{\partial F_k}{\partial f}\frac{\partial f}{\partial p_i} + \frac{\partial F_k}{\partial x}\frac{\partial x}{\partial p_i} + \frac{\partial F_k}{\partial p_i} = 0 \quad k = 1, 2$$

The previous system can also be written in the following matrix form, with J the Jacobian matrix.

$$J \cdot \begin{pmatrix} \partial f / \partial p_i \\ \partial x / \partial p_i \end{pmatrix} + \begin{pmatrix} \partial F_1 / \partial p_i \\ \partial F_2 / \partial p_i \end{pmatrix} = \begin{pmatrix} 0 \\ 0 \end{pmatrix}; \text{ hence}: \begin{pmatrix} \partial f / \partial p_i \\ \partial x / \partial p_i \end{pmatrix} = -J^{-1} \cdot \begin{pmatrix} \partial F_1 / \partial p_i \\ \partial F_2 / \partial p_i \end{pmatrix}$$

Only the first equation from the matrix relation is useful in this case, which finally yields:

$$\mathrm{Sp}_i = \frac{\partial f}{\partial p_i} = \frac{\frac{\partial F_1}{\partial x}\frac{\partial F_2}{\partial p_i} - \frac{\partial F_2}{\partial x}\frac{\partial F_1}{\partial p_i}}{\frac{\partial F_1}{\partial f}\frac{\partial F_2}{\partial x} - \frac{\partial F_1}{\partial x}\frac{\partial F_2}{\partial f}}$$

As a consequence, each absolute sensitivity can be expressed by means of 6 partial derivatives. The relative and normalized sensitivities derive from the absolute sensitivity.

### Sensitivity with respect to temperature

The characteristic coefficients of the sensor also depend on the temperature T. For each parameter $p_i$, this dependance is characterized by a linear temperature coefficient:

$$\tau_{p_i} = \frac{1}{p_i}\frac{\partial p_i}{\partial T}$$

In fact, for every material, two coefficients are specified by the manufacturer: a thermal expansion coefficient and a thermal stability coefficient of the permittivity. Then, the (relative) sensitivity of the frequency with respect to temperature is given by:

$$RS_T = \frac{1}{f}\frac{\partial f}{\partial T} = \sum_i \frac{1}{f}\frac{\partial f}{\partial p_i}\frac{\partial p_i}{\partial T} = \sum_i \frac{p_i}{f}\frac{\partial f}{\partial p_i}\tau_{p_i} = \sum_i \mathrm{NSp}_i\,\tau_{p_i}$$

It derives from the normalized sensitivities (the default ones in the package) and the linear temperature coefficients, so its computation is straightforward.

**Computational aspects**

As already mentioned, the computation of frequencies relies on a numerical technique, whereas the computation of sensitivities resorts to computer algebra capabilities. Each sensitivity is expressed by means of a symbolic formula including partial derivatives. These involve more or less intricate expressions, which are determined by a dynamic programming technique (a "memory function"), so the partial derivatives are evaluated only when needed and at most once per session. Then, they must be evaluated at a particular point, so the frequency must be computed jointly. Actually, the auxiliary function `Sensitivity` returns the (intricate) symbolic expression of the absolute sensitivity with respect to the specified parameter, and the function `NSensitivity` computes its (more useful) numerical value at the working frequency of the resonator.

> `Timing[NSensitivity[defaultSensor, h`$_3$` (* Type→Normalized *)]]`

> {0.4 Second, -19.3391}

> `Timing[NSensitivity[defaultSensor, h`$_3$`, Type → Relative]]`

> {0.333333 Second, -0.0580172}

> `Timing[NSensitivity[defaultSensor, h`$_3$`, Type → Absolute]]`

> {0.333333 Second, $-3.67881 \times 10^{10}$}

## 4. Design and use of the package

We are faced with an aspect of the general issue of representing more or less complicated engineering objects. This can be neglected when programs apply to mathematical problems of algorithmic nature. However, engineering applications of Mathematica involve data structures, which induce specific difficulties due to numerous parameters or possible evolutions of the representation.

**Representing a physical system**

From the point of view of a physicist or an engineer (who are assumed to be the potential users), there are two major levels of abstraction:
• the technical system with its geometrical and physical parameters, which as a rule have both names (symbols) and values. It is desirable to be able to switch over the nature (symbolic or numerical) of the parameters.
• the theoretical model giving the required (output) parameters from the data (input parameters). We must be aware that a single system may behave several models and each model may lead to different computational techniques.

From the computational point of view, these two levels are implemented by means of:
• a data structure (an expression) that describes the sensor with its parameters (symbols and values):

$$\text{Sensor[parameters]}$$

• the algorithms (the functions) drawn from the model for computing the output parameters (here, the resonant frequency and the sensitivities); choosing the model and the method is left to the user by means of options, with default values [8]; hence the schematic syntax:

$$\text{SomeOutputParameter[aSensor,Model→aModel,Method→aMethod]}$$

The use of options naturally encapsulates the representation of the model. Indeed, the "supplier" may add novel models or methods and possibly novel options without any change in the interface (the pattern), hence without disturbing the "customer".

## A data structure for sensors

Building a data structure with Mathematica consists in wrapping the elementary data in a list or an expression [9]. In this last case, the head plays the role both of a type and a constructor. In fact, the notion of a constructor comes from procedural languages, so that a separate constructor is most often useless in Mathematica, which uses the same syntax for data and functions. This functional approach assures the referential transparency of programs, for it avoids undesirable side effects due to the assignment of individual variables; it urges the user to process the data as a whole, by passing the entire structure as argument to functions.

However, specific problems emerge again in the case of engineering applications. When the arguments are numerous, it may be difficult for the user to remember their position; hence the usefulness of named parameters. Also, it should be possible to maintain and manipulate several sensor representations at the same time. Finally, the user may wish to choose (and possibly change) not only the values but also the symbols of the parameters, and possibly use different symbols for different sensors.

Hence the idea of a small database with three fields (attributes): the selectors, the symbols of the parameters and their values. Such a data structure allows referring to the parameters either by selectors or by names (symbols) and enables changing the names or the values. Finally, the wrapper Sensor[] types the list of parameters thus restricting its use to the particular context of sensor design.

```
Edit[defaultSensor, Style → Output]
```

| | | |
|---:|:---:|:---|
| SubstrateHeight | $h_1$ | $1.27 \times 10^{-3}$ |
| MedianLayerHeight | $h_2$ | $1.5 \times 10^{-2}$ |
| UpperLayerHeight | $h_3$ | $3. \times 10^{-3}$ |
| InnerRadius | $a$ | $1.75 \times 10^{-2}$ |
| OuterRadius | $b$ | $2.5 \times 10^{-2}$ |
| ResonatorRelativePermittivity | $\varepsilon_{dr}$ | $3.7 \times 10^{1}$ |
| SubstrateRelativePermittivity | $\varepsilon_1$ | $1.05 \times 10^{1}$ |
| MedianLayerRelativePermittivity | $\varepsilon_2$ | $1.$ |
| UpperLayerRelativePermittivity | $\varepsilon_3$ | $1.$ |

Actually, in order to avoid redundant information, the selectors are associated with the symbol `Sensor` as upvalues, so the sensor representation finally amounts to a table of parameter symbols and values. This table was implemented as a list of rules. The right-hand member of each rule can be a symbol, possibly the parameter symbol itself.

```
anIllustrativeSensor = Create[Sensor, Style -> Dialog]
```

Sensor[{$h_1 \to 0.00127$, $h_2 \to 0.015$, $h_3 \to h$, $a \to a$, $b \to b$, $\varepsilon_{rd} \to 37.$, $\varepsilon_1 \to 10.5$, $\varepsilon_2 \to 1.$, $\varepsilon_3 \to \varepsilon$}]

## Modularity and object-oriented features

In most cases, the functions `NFrequency` and `NSensitivity` will be called only interactively. However, the user may want to program complementary functions. So, we should foresee and facilitate possible future evolutions due to more sophisticated models or more detailed representations of objects. This requires a data encapsulation technique that hides internal transformations of the data structure. This is achieved by restricting access to data by means of selectors, which avoid direct interaction with the internal representation [10, 9].

Nevertheless, Mathematica is an open system so that every part (subexpression) of any object (expression) is accessible, except for a few atomic objects. As a consequence, encapsulation is a matter of programming style rather than a feature of the language; in other words, it is based on the user's responsability, who should accede to the data structure only by means of the selectors provided with in the package. Symmetrically, it is the programmer's responsibility for supplying a complete set of selectors.

```
ResonatorRelativePermittivity[defaultSensor]
```

{$\varepsilon_{dr} \to 37.$}

A generic selector that returns the values of the parameters specified by their symbols is included in the package. Its behavior is similar to that of `Options`.

```
Parameters[defaultSensor, {h₁, ε₁}]
```

{$h_1 \to 0.00127, \varepsilon_1 \to 10.5$}

The selectors and the other operators (e.g., `InnerRadius`, `NFrequency`) are associated with the symbol `Sensor` as upvalues, i.e., are methods in object oriented terms. This enables the possible use of the same generic commands for other types of objects. In particular, there is in the package a method called `New` that generates a default sensor with a list of default parameters (symbols and values); `New` returns a list of default symbolic parameters when used with the option `Symbolic` set to `True`. In the same way, `Create` directly edits a default sensor (see next section).

```
New[Sensor, Symbolic → True]
```

Sensor[{$h_1 \to h_1, h_2 \to h_2, h_3 \to h_3, a \to a, b \to b, \varepsilon_{dr} \to \varepsilon_{dr}, \varepsilon_1 \to \varepsilon_1, \varepsilon_2 \to \varepsilon_2, \varepsilon_3 \to \varepsilon_3$}]

From the object-oriented point of view, the head `Sensor` or more precisely the pattern `Sensor[{__Rule},___]` represents a class of objects, whereas any particular sensor is an instance of this class. The number of parameters or possible options may determine subclasses.

## Use of the package

This section can be viewed as a short user's guide for the package. Complementary information can be found in the notebook "SensorDesign.nb".

In the current model, there are nine geometrical and physical parameters, given as a list of rules that specify their symbols and values. The command `Create` puts into an input cell a `GridBox` were the user can enter these parameters. It contains default parameter symbols and values. Thanks to a `TagBox` hidden inside the underlying expression, it is directly interpreted as a sensor when evaluated. Removing a value yields a parameter with a symbolic value ; in the example below, the first three values were removed and the next two parameter symbols modified.

A user may wish to modify a previously defined sensor. This can be done with the command `Edit`, similar to `Create`, which builds an input cell with a `GridBox` where the user can modify the parameters. The same command outputs a 2-D tabular form of the parameters when used with the option `Style` set to `Output` (see above).

```
Edit[defaultSensor]
```

- Cell -

| | | |
|---:|:---:|:---|
| **SubstrateHeight** | $h_1$ | ☐ |
| **MedianLayerHeight** | $h_2$ | ☐ |
| **UpperLayerHeight** | $h_3$ | ☐ |
| **InnerRadius** | r | $1.75 \times 10^{-2}$ |
| **OuterRadius** | R | $2.5 \times 10^{-2}$ |
| **ResonatorRelativePermittivity** | $\varepsilon_{dr}$ | $3.7 \times 10^1$ |
| **SubstrateRelativePermittivity** | $\varepsilon_1$ | $1.05 \times 10^1$ |
| **MedianLayerRelativePermittivity** | $\varepsilon_2$ | 1. |
| **UpperLayerRelativePermittivity** | $\varepsilon_3$ | 1. |

Sensor[{$h_1 \to h_1, h_2 \to h_2, h_3 \to h_3, r \to 0.0175, R \to 0.025, \varepsilon_{dr} \to 37., \varepsilon_1 \to 10.5, \varepsilon_2 \to 1., \varepsilon_3 \to 1.$}]

This feature works only with version 3 (or higher). For users of older versions, the functions `ChangeParameterSymbols`, `ChangeParameterValues` and `ClearParameters` have a similar effect.

Finally, the functions `NFrequency` and `NSensitivity` have two options for choosing the model (`ElectricWall` or `MagneticWall`) and the numerical method (`NewtonMethod` or `SecantMethod`). Their default values are `MagneticWall` for the former and `SecantMethod` for the latter.

## 5. Comparison with experimental measurements

The model is validated by comparison with experimental data. In the following, theoretical results will be in solid lines and experimental data in dotted lines. The latter are read from the file "ExperimentalData.txt" where the values of the parameters and the experimental data are stored directly as Mathematica expressions.

```
Shallow[theseData = << "ExperimentalData.txt"]
```

{ParameterValues → { ≪9≫ }, Frequencies → { ≪25≫ }, Units → { ≪2≫ }}

```
Units /. theseData
```

{Meter Milli, Giga Hertz}

These data correspond to the resonant frequencies of the $TE_{01\delta}$ mode (in GHz) versus $h_3$ (mm). The remaining parameter numerical values are associated with the data.

```
s[η_] := ChangeParameterValues[Sensor[ParameterValues /. theseData], {h[3] → η * 10.^-3}]
```

Now, we can compare the measured frequencies with the theoretical ones, computed by means of a magnetic-wall model, which is more suitable when b/a is large.

```
g₀ = ListPlot[Frequencies /. theseData,
    PlotStyle → {PointSize[0.01]}, DisplayFunction → Identity];
g₁ = Plot[NFrequency[s[η]] / 10^9, {η, 0, 14}, PlotPoints → 5,
    PlotDivision → 1, DisplayFunction → Identity];
Show[g₀, g₁, AxesLabel → {"h₃ (mm)", "f(GHz)"}, AxesOrigin → {0, 1.06},
 DisplayFunction → $DisplayFunction]
```

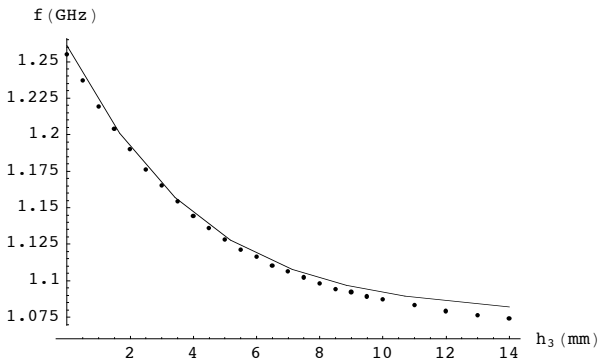

- Graphics -

Then, the sensitivities with respect to $h_3$ can be computed from the previous experimental data and compared with the theoretical ones. Below, we fit the data to a polynomial, then compute the derivative of the resulting function. The coefficient $10^3$ is necessary in $g_2$ because $\eta = 10^{-3} \, h_3$.

```
S_h[3] [η_] = Fit[Frequencies /. theseData, η^Range[0, 4], η];
g₂ = Plot[Evaluate[10³ D[S_h[3][η], η]], {η, 0, 14},
    PlotStyle → Dashing[{0.03, 0.03}], DisplayFunction → Identity];
g₃ = Plot[10⁻⁹ (NSensitivity[s[η], h[3], Type → Absolute]), {η, 0, 14},
    PlotPoints → 5, PlotDivision → 1, DisplayFunction → Identity];
Show[g₂, g₃, DisplayFunction → $DisplayFunction,
 AxesLabel → {"h₃ (mm)", "S_h₃ (KHz/μm)"}, AxesOrigin → {0, -45}]
```

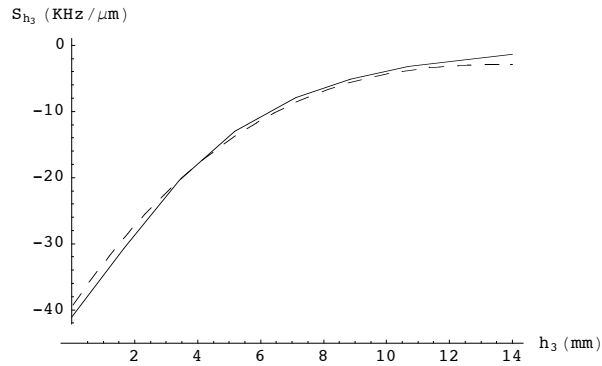

- Graphics -

## 6. Conclusion

A simple analytical model based on the transverse resonance method can be used to evaluate the resonant frequency of cylindrical multilayer dielectric sensors. This model also enables the sensitivity analysis with respect to geometrical or physical parameters. It is validated by comparison with experimental data. The $TE_{01\delta}$ and $TM_{01\delta}$ modes are computed with a precision of about 1 percent (10 percent in the worst case for the $TM_{01\delta}$ mode).

A package was designed to implement the functions that determine the frequency and the sensitivities from this model. The computation of the frequency is based on a numerical approximation, whereas the sensitivity is obtained from a combination of symbolic and numerical computations. Besides, we tried to bring out a few design patterns for the computational description of a technical system.

The model and the corresponding computations were useful to understand the general properties of shielded cylindrical multilayer dielectric resonators when used as resonant transducers. They brought to the fore the role of geometrical and physical parameters. So, the package can be used for the computer assisted design of this range of sensors. Possible applications range from displacement, thickness or pressure sensors to humidity probes or devices for measuring the electromagnetic parameters of materials.

### Acknowledgements

The authors wish to thank their colleague Pierre Geveaux for his help to work out the package.

### Additional Material

The package "SensorDesign.m", the notebooks "SensorModelling.nb" and "SensorDesign.nb" (documentation), and the text file "ExperimentalData.txt" are available at the URL http://macmaths.ens2m.fr/Mathematica/packages.